\documentclass[conference, a4paper]{IEEEtran}
\usepackage{amsthm}

\usepackage{enumitem}
\newcounter{problem}
\newcounter{subproblem}[problem]

\usepackage{balance}

\usepackage{hyperref}

\usepackage{mathtools}

\usepackage{color}
\usepackage{xcolor}
\usepackage{colortbl}
\definecolor{purple}{RGB}{139, 0, 139}
\usepackage{manfnt}
\usepackage{cleveref}
\usepackage{url}

\newif\iftodo   
\todotrue
\newif\iftodoshort  
\todoshortfalse
\ifCLASSINFOpdf
  \usepackage[pdftex]{graphicx}
   \DeclareGraphicsExtensions{.pdf,.jpeg,.png,.tiff}
\else
   \usepackage[dvips]{graphicx}
   \DeclareGraphicsExtensions{.eps,.ps}
\fi
\usepackage{amssymb,amsmath}
\usepackage[mathscr]{euscript}
\usepackage{bm} 
\usepackage{bbm}
\usepackage{dsfont}

\usepackage{algorithmic}
\usepackage[ruled,vlined,commentsnumbered]{algorithm2e}

\graphicspath{ {./fig/} }
\newcommand{\Rmnum}[1]{\uppercase\expandafter{\romannumeral #1}}
\newcommand{\rmnum}[1]{\lowercase\expandafter{\romannumeral #1}}

\usepackage[utf8]{inputenc}

\usepackage{hyperref}
\usepackage{amsthm}
\usepackage{amssymb,amsmath}
\usepackage[mathscr]{euscript}

\newcommand{\cosl}[1]{}
\newcommand{\resl}[1]{}

\usepackage[draft,footnote,nomargin]{fixme}
\newcommand{\fnql}[1]{}
\newcommand{\fnsv}[1]{}

\IEEEoverridecommandlockouts

\begin{document}

%
\title{Downlink Performance of Uplink Fractional Power Control in 5G Massive MIMO Systems}
\author{
\IEEEauthorblockN{Paolo Baracca\IEEEauthorrefmark{1}, Lorenzo Galati Giordano\IEEEauthorrefmark{2}, Adrian Garcia-Rodriguez\IEEEauthorrefmark{2},\\ Giovanni Geraci\IEEEauthorrefmark{3}, and David L\'{o}pez-P\'{e}rez\IEEEauthorrefmark{2}}
\IEEEauthorblockA{\IEEEauthorrefmark{1}Nokia Bell Labs, Stuttgart, Germany\\ 
\IEEEauthorrefmark{2}Nokia Bell Labs, Dublin, Ireland\\
\IEEEauthorrefmark{3}Universitat Pompeu Fabra, Barcelona, Spain}
}

\maketitle
\begin{abstract}
Uplink power control is an efficient scheme to mitigate pilot contamination in massive multiple-input multiple-output (MIMO) systems.
In this work, we provide a comprehensive study on the effects of fractional power control (FPC) on the downlink performance of the most relevant fifth generation (5G) massive MIMO deployments.
Specifically, we perform thorough system simulations based on the most recent three dimensional spatial channel model released by the 3rd Generation Partnership Project to evaluate the impact of different deployment-related parameters such as pilot reuse factor, beamforming criterion, and base station array size.
Our results indicate the most suitable tuning of the FPC parameters and show that optimized FPC provides huge gains in the cell border throughput when compared to a baseline scheme with all the users transmitting at maximum power.
Moreover, our simulations also demonstrate that the effectiveness of FPC grows in scenarios with severe pilot contamination, confirming that implementing this feature is essential in realistic deployments.
\end{abstract}

\section{Introduction}

Massive multiple-input multiple-output (MIMO) is a fundamental enabler in the fifth generation (5G) of cellular networks to cope with the increasing demand for higher data throughput \cite{boccardi_cm14}.
Base stations (BSs) equipped with tens to hundreds of antennas provide much higher spectral efficiency when compared to the previous generation BSs, thanks to two complementary techniques: beamforming and spatial multiplexing \cite{larsson_cm14}.
With beamforming, the signals sent by the multiple BS antennas add up constructively at the receiver, thus boosting the received useful signal power.
In addition to that, multiple streams can be sent simultaneously by the BS via spatial multiplexing, with these streams separated in the spatial domain by using proper precoders.
An intense massive MIMO research activity has been performed, both in the academia and in the industry, in the last few years, and, thanks to that, some basic schemes like the full-dimension MIMO have already been standardized by the 3rd Generation Partnership Project (3GPP) \cite{nam_cm12}. 

One of the key operations of massive MIMO systems is the channel state information (CSI) acquisition at the BS. In frequency division duplex (FDD), channels are first estimated at the user equipment (UE) devices, and then provided back to the BS via feedback, whereas, in time division duplex (TDD), channel reciprocity is exploited, and the UEs transmit pilot sequences to allow channel estimation directly at the BSs.
As the training overhead is proportional to the number of BS antennas in FDD, the initial massive MIMO deployments are expected to be in TDD mode \cite{bjoernson_cm16}.
Due to the limited resources available for training in massive MIMO TDD, the pilot sequences are reused by UEs of different cells, and that causes pilot contamination, which has been recognized as one of the most limiting factors in massive MIMO systems \cite{hoydis_jsac13}.

Several works have been carried out in the recent years to mitigate the effects of pilot contamination \cite{elijah_survey16}, and one of the most practical and promising solutions seems to be a right mixture of pilot coordination among the cells \cite{galati_wcnc18} and uplink power control (PC) \cite{saxena_vtc15}.
Note that uplink PC has been used as a technique to limit the uplink interference also in the Long Term Evolution (LTE) standard \cite{castellanos_vtc08}.
In line with this consideration, several techniques have been recently proposed.
In \cite{guo_icc14}, a lower bound on the achievable uplink signal to interference plus noise ratio (SINR) is derived, and, based on that, uplink pilot and data powers are optimized to maximize the energy efficiency.
This optimization is further decentralized by using game theory in \cite{guo_ew15}.
A joint optimization of pilot coordination among cells and uplink PC is performed in \cite{vanchien_twc18}, where, differently from the previous works, the pilot signals are modeled as continuous optimization variables.
A similar joint optimization, but considering both uplink and downlink performance is carried out in \cite{filho_telt17}.
Moreover, the benefits of uplink PC in massive MIMO systems have been evaluated in \cite{saxena_vtc15}, and indeed some algorithms have been tested in some recent field trials \cite{hasan_vtc17}.

In this work, we focus on maximizing the downlink system performance of massive MIMO TDD systems by applying open-loop uplink fractional power control (FPC) to the UE-generated pilots.
Differently from most of the previous works that assume a setup with simplified assumptions like Rayleigh fading \cite{guo_icc14, guo_ew15, vanchien_twc18, filho_telt17}, our main objective is to provide a thorough understanding of the performance of FPC in realistic 5G massive MIMO settings.
We do this by relying on the accurate three dimensional (3D) spatial channel model developed by 3GPP \cite{3gpp_tr38901}.
Motivated by the promising results of \cite{saxena_vtc15}, we perform extensive system simulations by considering different configurations in terms of pilot reuse scheme, beamforming criterion and BS array size in order to understand: a) how FPC should be optimized and b) to which extent FPC is capable to mitigate pilot contamination.
Our numerical results corroborate that an optimized FPC provides significant gains in practical 5G scenarios, mainly for the cell-edge UEs, when compared to a baseline scheme where each UE just transmits at maximum power its assigned pilot sequence.


\section{System Setup}

In this work, we consider the 3D spatial channel model proposed by 3GPP \cite{3gpp_tr38901}.
We focus on the urban macro (UMa) scenario where each BS is at 25 m height, and assume a hexagonal deployment with nineteen sites, an inter-site distance of 500 m, three sectors per site and wraparound.
Each BS transmits with a maximum power of 46 dBm over a system bandwidth of 10 MHz operating at a carrier frequency of 2 GHz: we assume the band to be divided into $N=50$ resource blocks (RBs) with a bandwidth of 180 kHz each.
The BSs are equipped with $M$ cross-polarized antennas, organized in a uniform planar array with 4 rows and $M/4$ columns,  with a mechanical downtilt of 12$^{\circ}$ and an antenna spacing of $0.5\lambda$, where $\lambda$ is the carrier wavelength.
Each antenna element has a parabolic radiation pattern with 8 dBi gain, 65$^{\circ}$ half-power beam-width (HPBW) and a front-to-back ratio (FBR) attenuation of 30 dB. Single-antenna UEs are randomly distributed in the network, with $20\%$ of them being outdoor and $80\%$ indoor in buildings, whose height is uniformly distributed between 4 and 8 floors and where the floor height is 3 m.
A summary of the system simulation parameters can be found in Tab. \ref{tab_parameters} and more details in \cite{3gpp_tr38901}.

\begin{table}
\centering
\caption{System simulation parameters.}
\label{tab_parameters}
\begin{tabular}{|l|p{4cm}|}
\hline
    \textbf{Parameter} 			& \textbf{Description} \\ \hline
    BS deployment				& Hexagonal with wrap-around, 19 sites, 3 sectors per site \\ \hline
   	Carrier frequency 			& 2 GHz \\ \hline
	System bandwidth 			& 10 MHz \\ \hline
	Duplexing mode				& TDD \\ \hline
	Channel model				& 3GPP 3D UMa \cite{3gpp_tr38901}  \\ \hline
    BS inter-site distance 	& 500 m \\ \hline
    BS height 					& 25 m \\ \hline    
	UE distribution				& Uniform \\ \hline
	BS max transmit power		& 46 dBm \\ \hline
	BS antenna array			& Uniform planar array of $4 \times M/4$ cross-polarized elements \\ \hline
	BS antenna array downtilt	& $12^{\circ}$ \\ \hline
	BS antenna element spacing	& $0.5\lambda$ \\ \hline
	BS antenna element gain 	& 8 dBi \\ \hline
	BS antenna element HPBW 	& $65^{\circ}$ \\ \hline
	BS antenna element FBR 		& 30 dB \\ \hline
	UE max transmit power		& 23 dBm \\ \hline
	UE antenna array			& Single omni-directional antenna element \\ \hline
	Traffic model				& Full buffer \\ \hline
\end{tabular}
\end{table}

\subsection{Subframe Structure and Channel Estimation}

In this work, we concentrate on the downlink performance of a fully loaded system. In this setup, UEs transmit uplink pilot sequences to allow channel estimation at the BSs.
We assume that transmission is organized in subframes, each made by $T=14$ orthogonal frequency division multiplexing (OFDM) symbols, and lasting 1 ms \cite{galati_wcnc18}.
In the first $\tau$ OFDM symbols of each subframe each scheduled UE $k$ transmits its assigned pilot sequence with power $P_k$.
Subsequently, each BS computes the beamformers based on the least-squares estimate of the channels connecting itself to its scheduled UEs.
The remaining $T-\tau$ OFDM symbols are allocated to the downlink data transmission performed by the BSs.

We assume that $K=16$ UEs are scheduled by each BS on each subframe, and compare two reuse factors in the pilot sequence allocation to the UEs \cite{galati_wcnc18}.
\begin{itemize}
\item With reuse 1 (R1), we have $\tau=1$, and the same set of pilot sequences is reused by all the sectors in the network.
Although this pattern requires just about $7\%$ overhead for channel training, it introduces strong pilot contamination in the system.
\item With reuse 3 (R3), we have $\tau=3$, and thus orthogonal pilot sequences are allocated among the three sectors of the same site. This last scheme increases the training overhead to about $21\%$ in an attempt to limit the impact of pilot contamination.
\end{itemize}

\subsection{Downlink Transmission}

Let us denote with the row vector $\bm{h}_{k,j,n}$, $k=1,2,\ldots,K$, the channel estimated at BS $j$ for its scheduled UE $k$ on RB $n$.
We assume that all the $K$ UEs of a sector are scheduled on all the RBs, and two beamforming criteria are compared:
\begin{itemize}
\item Maximum ratio transmission (MRT): the beamformer is designed for each UE to maximize the signal to noise ratio (SNR) at that UE \cite{lo_icc99};
\item Zero forcing (ZF): the beamformer is designed for each UE to minimize the interference generated toward the co-scheduled UEs \cite{spencer_tsp04}.
\end{itemize}
In more detail, after defining with $\bm{H}_{j,n}=\left[ \bm{h}_{1,j,n}^T, \bm{h}_{2,j,n}^T, \ldots, \bm{h}_{K,j,n}^T, \right]^T$ the resultant channel matrix estimated at BS $j$ on RB $n$, the beamformers used for downlink transmission can be written as
\begin{equation}
\bm{W}_{j}^{\rm (MRT)} = \bm{H}_{j,n}^H \bm{D}_{j,n}^{\rm (MRT)} \,,
\end{equation}
\begin{equation}
\bm{W}_{j}^{\rm (ZF)} = \bm{H}_{j,n}^H \left( \bm{H}_{j,n} \bm{H}_{j,n}^H \right)^{-1} \bm{D}_{j,n}^{\rm (ZF)} \,,
\end{equation}
where the $K \times K$ diagonal matrices $\bm{D}_{j,n}^{\rm (MRT)}$ and $\bm{D}_{j,n}^{\rm (ZF)}$ are computed to meet the power constraint at the BS under the assumption of equal downlink power allocation among the UEs.

\section{Fractional Uplink Power Control}

By allowing the UEs at the cell center to transmit at lower power than the UEs at the cell edge, uplink PC has been recognized since the start of cellular networks as an efficient way to manage intra-cell interference and balance the different SINR conditions of the UEs \cite{zander_tvt92}.
Following the current 5G standardization direction, we focus in this work on the FPC mechanism already developed for LTE, and numerically optimize its parameters as a function of different massive MIMO configurations.
Our aim is to verify its effects on the downlink system performance, with particular attention at the cell edge UEs.

With open-loop FPC, the power $P_k$ used by UE $k$ to transmit its pilot sequence can be written, in logarithmic scale, as \cite{castellanos_vtc08}
\begin{equation}
P_k = \min \left\{ P^{\rm (UE)} , \, P_0 + 10\log_{10}(N) +\alpha L_k  \right\}\,,
\label{eq_fpc}
\end{equation}
where $P^{\rm (UE)}=23$ dBm is the maximum transmit power at the UE, $L_k$ is the large scale fading attenuation between UE $k$ and its anchor BS (which includes path-loss, shadowing and antenna element gain), $P_0$ is a parameter used to control the per-RB SNR target, and $\alpha\in[0,1]$ defines the fractional compensation factor of the large scale fading attenuation.
By properly setting the two parameters $P_0$ and $\alpha$ in (\ref{eq_fpc}), different working modes can be obtained.
\begin{itemize}
\item $\alpha=0$: In this case, we impose that all the UEs transmit at the same power, and the UE transmit power is mainly regulated by the value of $P_0$. In the following, as a baseline case, we will force all the UEs to transmit at full power in (\ref{eq_fpc}), i.e., $P_k=P^{\rm (UE)},\,\forall k$, and refer to this case as no power control (noPC). 
\item $\alpha=1$: In this case, FPC tries to fully compensate the large scale fading attenuation. Because of the constraint on the maximum transmit power $P^{\rm (UE)}$, this policy is such that some UEs will scale down their transmit power to achieve a certain SNR target, which depends on $P_0$, whereas the remaining UEs will all transmit at maximum power. One of the limits of this configuration is that, if the value of $P_0$ is too high, a large number of UEs will transmit at maximum power generating a high level of interference and pilot contamination.
\item $0 < \alpha <  1$: This corresponds to the real application of FPC. A higher value of $\alpha$ introduces less difference in the uplink SNR, i.e., it guarantees more fairness among the UEs, but needs to be coupled with a lower value of $P_0$ in order to avoid a strong level of interference in the network.
\end{itemize}

\section{Performance Evaluation}
\label{sect_results}

The main objective of this work is to numerically optimize the system parameters $P_0$ and $\alpha$ in (\ref{eq_fpc}) in realistic scenarios for the most important massive MIMO deployments.

\begin{figure}[t!]
\centering
\includegraphics[width=1\hsize]{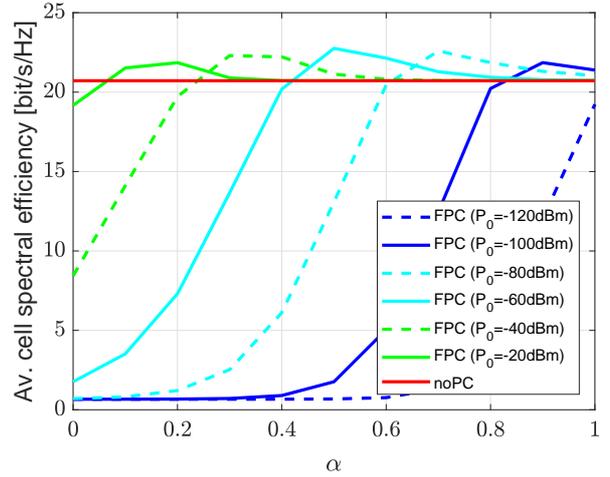}\\
\caption{Average cell spectral efficiency for different values of $P_0$ and $\alpha$ with $M=128$, ZF and R3.}
\label{cse_versus_alpha_ZF_128A_R3}
\end{figure}

\begin{figure}[t!]
\centering
\includegraphics[width=1\hsize]{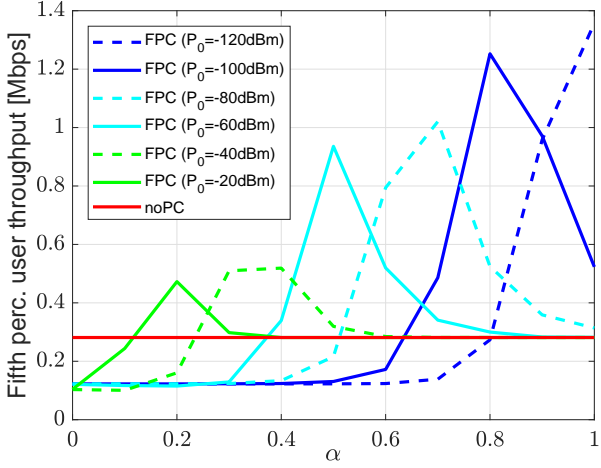}\\
\caption{Fifth percentile of the UE throughput for different values of $P_0$ and $\alpha$ with $M=128$, ZF and R3.}
\label{cbt_versus_alpha_ZF_128A_R3}
\end{figure}

In Fig. \ref{cse_versus_alpha_ZF_128A_R3} and Fig. \ref{cbt_versus_alpha_ZF_128A_R3}, we assume $M=128$, reuse 3 and ZF, and report the average cell spectral efficiency (CSE) and the fifth percentile of the UE throughput, respectively, versus $\alpha$ for different values of $P_0$.
Note that these two key performance indicators (KPIs) are the most commonly used by 3GPP, with the fifth percentile of the UE throughput often being indicated also as cell border throughput (CBT).
In these figures, we compare FPC against the baseline noPC, where all the UEs transmit at maximum power. 
First, we observe that, depending on the value of $P_0$, there is an optimal value of $\alpha$ that maximizes each KPI.
For instance, with $P_0=-60$ dBm, $\alpha=0.5$ maximizes both the CSE and the CBT, whereas with $P_0=-100$ dBm, $\alpha=0.8$ maximizes the CBT and $\alpha=0.9$ the CSE.
When looking at the absolute values, and when compared to noPC, we observe that FPC provides a limited gain of up to $10\%$ in the CSE, but a much higher gain of up to $350\%$ in the CBT, confirming that an optimized FPC is able to strongly improve the performance of the UEs severely limited by pilot contamination.
Moreover, we observe that, to maximize the CSE, a good solution is obtained by selecting $P_0=-60$ dBm and $\alpha=0.5$.
In contrast, to maximize the CBT, a better solution consists in selecting a lower value of $P_0$ and a higher value of $\alpha$.
In fact, although the CBT is maximized with $P_0=-120$ dBm and $\alpha=1$, i.e., with full compensation of the large scale fading attenuation, this specific FPC configuration is outperformed by the baseline noPC in the CSE.
This suggests that a better solution for the CBT, which also takes the CSE performance into account, consists in choosing $P_0=-100$ dBm and $\alpha=0.8$.
For the sake of clarity, in the following, we define:
\begin{itemize}
\item FPC$_{0.5}$ to denote the configuration with $P_0=-60$ dBm and $\alpha=0.5$ that maximizes the CSE;
\item FPC$_{0.8}$ to denote the configuration with $P_0=-100$ dBm and $\alpha=0.8$ that is more favorable for the CBT.
\end{itemize}

\begin{figure}[t!]
\centering
\includegraphics[width=1\hsize]{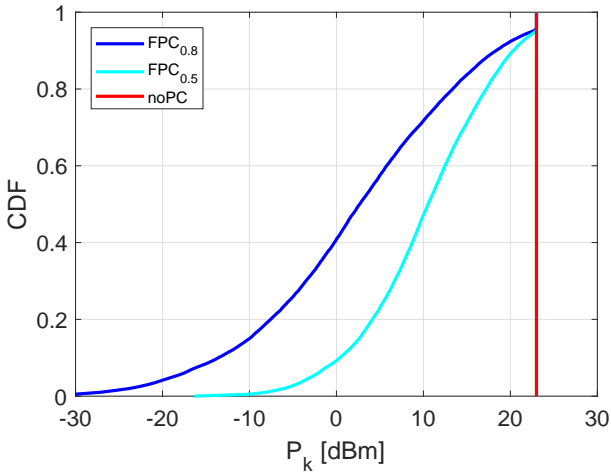}\\
\caption{CDF of the UE transmit power $P_k$ with $M=128$ and R3.}
\label{cdf_p_ZF_128A_R3}
\end{figure}

\begin{figure}[t!]
\centering
\includegraphics[width=1\hsize]{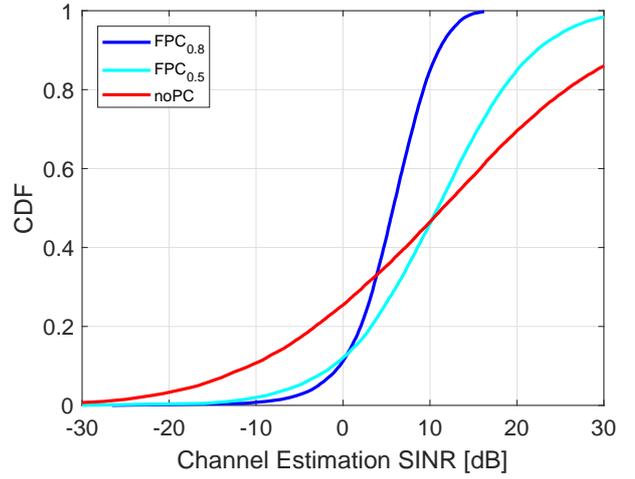}\\
\caption{CDF of the channel estimation SINR with $M=128$ and R3.}
\label{cdf_sinr_ZF_128A_R3}
\end{figure}

To better understand the previous results, we show in Fig. \ref{cdf_p_ZF_128A_R3} and Fig. \ref{cdf_sinr_ZF_128A_R3} the cumulative distribution function (CDF) of the UE transmit power $P_k$ and of the resulting channel estimation SINR at the BS, respectively, for the same setup.
We observe in Fig. \ref{cdf_p_ZF_128A_R3} that, with both FPC configurations, just about $5\%$ of the UEs transmit at full power, with these UEs being the ones experiencing the worst SNR conditions.
All the remaining UEs scale their transmit power down, with a higher transmit power reduction as expected in FPC$_{0.8}$ because of the lower value of $P_0$.
As a consequence, in Fig. \ref{cdf_sinr_ZF_128A_R3}, FPC allows to strongly reduce pilot contamination, thus improving the channel estimation SINR at the BS and the fairness among the UEs. 
When compared to noPC, we observe a gain in the fifth percentile of the channel estimation SINR of about 12 dB and 14 dB achieved by FPC$_{0.5}$ and FPC$_{0.8}$, respectively.

\begin{figure}[t!]
\centering
\includegraphics[width=1\hsize]{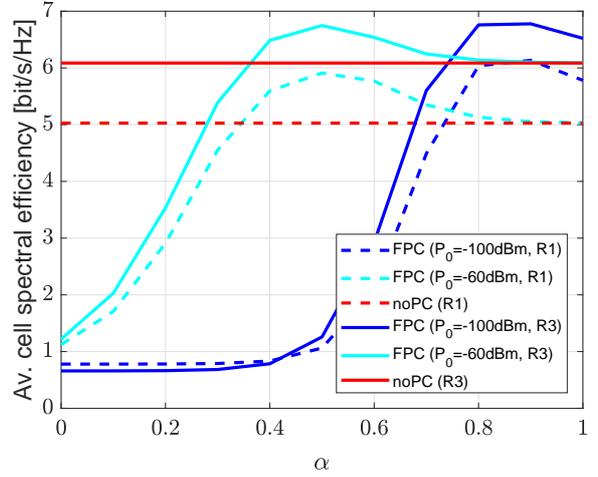}\\
\caption{Average cell spectral efficiency for different values of $P_0$ and $\alpha$ with $M=64$, MRT and comparing R1 against R3.}
\label{cse_versus_alpha_MF_64A}
\end{figure}

\begin{figure}[t!]
\centering
\includegraphics[width=1\hsize]{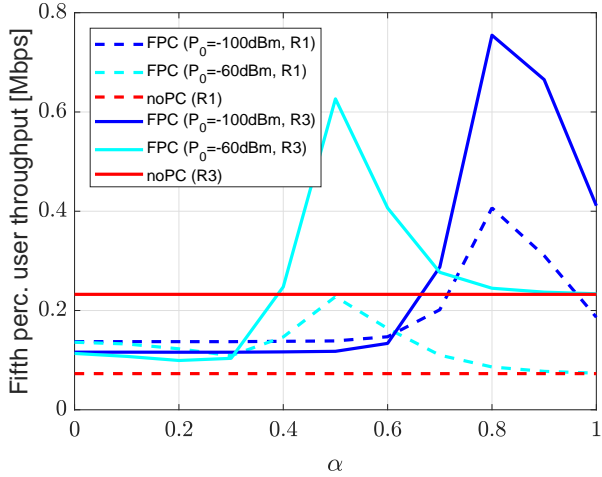}\\
\caption{Fifth percentile of the UE throughput for different values of $P_0$ and $\alpha$ with $M=64$, MRT and comparing R1 against R3.}
\label{cbt_versus_alpha_MF_64A}
\end{figure}

In Fig. \ref{cse_versus_alpha_MF_64A} and Fig. \ref{cbt_versus_alpha_MF_64A}, we report the average CSE and CBT versus $\alpha$ in a different setup, which considers $M=64$, MRT, and both R1 and R3.
First, in line with Fig. \ref{cse_versus_alpha_ZF_128A_R3} and Fig. \ref{cbt_versus_alpha_ZF_128A_R3}, we observe that $\alpha=0.5$ is a good choice for $P_0=-60$ dBm and $\alpha=0.8$ for $P_0=-100$ dBm.
Moreover, we observe that the gain achieved by FPC when compared to the baseline noPC is higher with R1 than with R3. For instance, FPC$_{0.8}$ provides a gain in the CBT that increases from about $230\%$ with R3 to about $450\%$ with R1.
This is because pilot contamination is limiting more the system performance with R1 than with R3, and, as a consequence, a scheme like FPC, specifically reducing pilot contamination, achieves more gain in the scenario with R1.

\begin{figure}[t!]
\centering
\includegraphics[width=1\hsize]{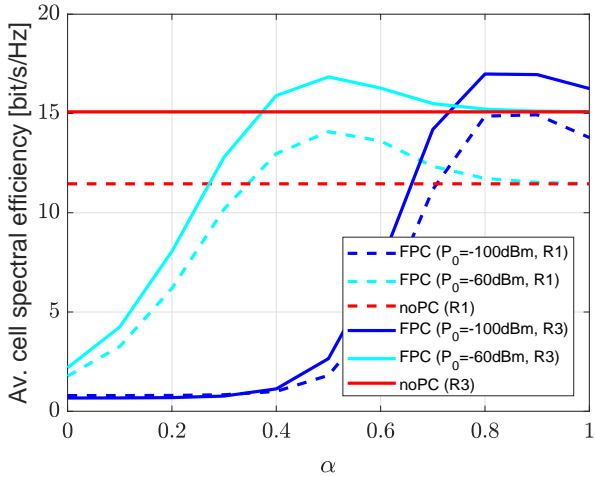}\\
\caption{Average cell spectral efficiency for different values of $P_0$ and $\alpha$ with $M=256$, MRT and comparing R1 against R3.}
\label{cse_versus_alpha_MF_256A}
\end{figure}

\begin{figure}[t!]
\centering
\includegraphics[width=1\hsize]{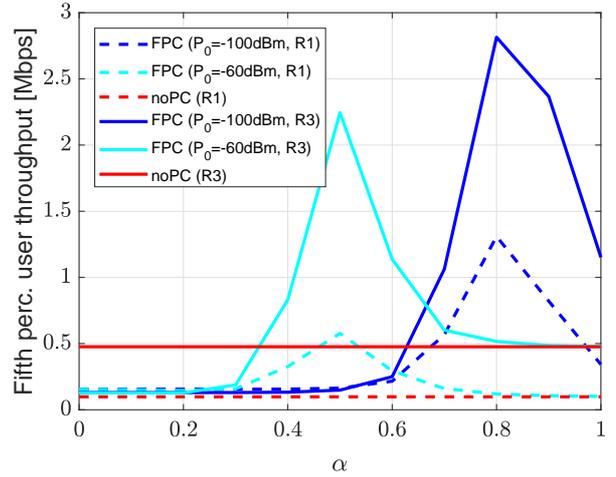}\\
\caption{Fifth percentile of the UE throughput for different values of $P_0$ and $\alpha$ with $M=256$, MRT and comparing R1 against R3.}
\label{cbt_versus_alpha_MF_256A}
\end{figure}

Then, similar results are observed in Fig. \ref{cse_versus_alpha_MF_256A} and Fig. \ref{cbt_versus_alpha_MF_256A}, where we increase the number of BS antennas to $M=256$.
These additional results allow us to conclude that, for a given $P_0$, the value of $\alpha$ that maximizes each KPI is quite constant, and does not depend much on the BS array size, the reuse factor and the beamforming criterion.
Moreover, when comparing for instance, Fig. \ref{cbt_versus_alpha_MF_64A} and Fig. \ref{cbt_versus_alpha_MF_256A}, we also observe that FPC achieves larger gains in the CBT when the number of BS antennas increases. Indeed, the gain achieved by FPC$_{0.8}$ when compared to noPC goes with R3 from about $230\%$ with $M=64$ to about $500\%$ with $M=256$.

\begin{figure}[t!]
\centering
\includegraphics[width=1\hsize]{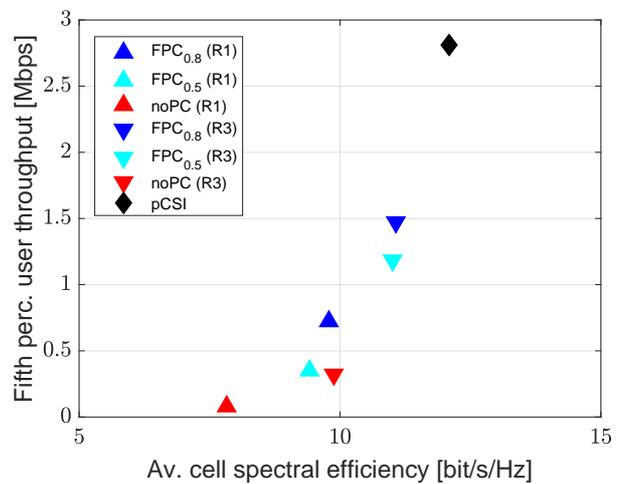}\\
\caption{Fifth percentile of the UE throughput versus average cell spectral efficiency with $M=128$ and MRT.}
\label{cbt_vs_cse_MF_128A}
\end{figure}

\begin{figure}[t!]
\centering
\includegraphics[width=1\hsize]{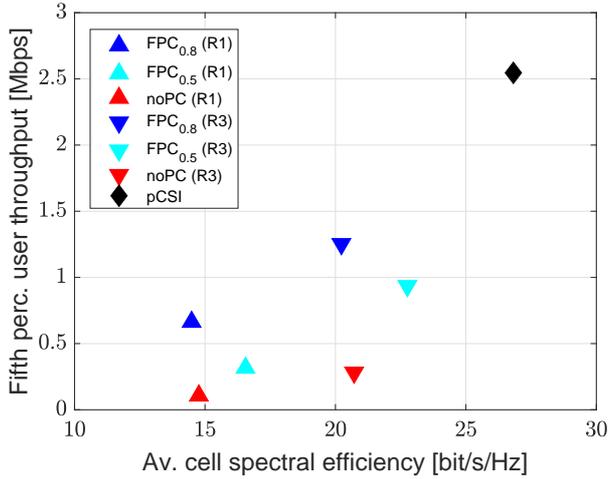}\\
\caption{Fifth percentile of the UE throughput versus average cell spectral efficiency with $M=128$ and ZF.}
\label{cbt_vs_cse_ZF_128A}
\end{figure}

Finally, in Fig. \ref{cbt_vs_cse_MF_128A} and Fig. \ref{cbt_vs_cse_ZF_128A}, we assume $M=128$, and compare, for MRT and ZF, respectively, the performance achieved by FPC against an upper bound obtained by assuming perfect CSI (pCSI), i.e., with neither noise nor pilot contamination at the BS.
Note that, for a fairer comparison, we also take into account the R3 training overhead of about $21\%$ for the pCSI performance.
First, we observe that slightly higher gains are obtained by FPC when compared to noPC with MRT than with ZF.
Then, more importantly, FPC with R1 allows achieving a similar performance to noPC with R3. For instance, with MRT in Fig. \ref{cbt_vs_cse_MF_128A}, we observe that FPC with R1 achieves almost the same CSE of noPC with R3, while providing similar or even better CBT.
This is an important result as FPC is already implemented and standardized, while advanced pilot reuse schemes require further implementation and development to allow coordination among the BSs for the pilot allocation.
Eventually, it is important to note that even with an optimized FPC, we can achieve about $90\%$ of the CSE but just $50\%$ of the CBT with respect to the pCSI case, showing that there is still room for additional mechanisms to further mitigate pilot contamination.

\section{Conclusions}

In this paper, we have performed a thorough evaluation of the benefits of uplink FPC to mitigate pilot contamination in the downlink of massive MIMO systems.
In order to provide an exhaustive understanding of the achievable performance with FPC in realistic 5G settings, we have carried out accurate system simulations in the most relevant massive MIMO deployments by considering a variety of pilot reuse factors, beamforming criteria, and BS array sizes.
The results of our work illustrate that designing an optimized FPC is fundamental for improving system performance, in particular the cell border throughput.
We have shown that the optimal value of $\alpha$ in terms of downlink performance is quite robust, i.e., independent of the pilot reuse factor, beamforming criterion, number of BS antennas, and KPI. Moreover, we have concluded that the configuration with $P_0=-60$ dBm and $\alpha=0.5$ represents a good solution for the cell spectral efficiency, whereas better performance in the cell border throughput can be obtained with $P_0=-100$ dBm and $\alpha=0.8$.
Finally, we have observed that FPC becomes essential in interference limited systems.
Indeed, we have demonstrated that the performance gains of FPC: a) increase when the number of BS antennas grows large, b) are typically higher with MRT when compared to ZF, and c) are higher with reuse 1 when compared to reuse 3.

\section{Acknowledgment}

This work has been performed in the framework of the Horizon 2020 project ONE5G (ICT-760809) receiving funds from the European Union. The authors would like to acknowledge the contributions of their colleagues in the project, although the views expressed in this contribution are those of the authors and do not necessarily represent the project.

\balance

\bibliographystyle{IEEEtran}
\bibliography{IEEEabrv,full_bibliography}

\end{document}